\documentclass[12pt]{iopart}
\usepackage{cite}
\usepackage{graphicx}
\begin{document}

\title[TRANSFER MATRIX METHOD FOR TRANSPORT IN GRAPHENE]{A Transfer Matrix Approach to Electron Transport in Graphene through Arbitrary Electric and Magnetic Potential Barriers}


\author{Sameer Grover$^1$, Sankalpa Ghosh$^1$ and Manish Sharma$^2$}
\address{$^1$ Department of Physics, Indian Institute of Technology Delhi, Hauz Khas, New Delhi, India}
\address{$^2$ Centre for Applied Research in Electronics, Indian Institute of Technology Delhi, Hauz Khas, New Delhi, India}

\begin{abstract}
A transfer matrix method is presented for 
solving the scattering problem for the quasi one-dimensional  
massless Dirac equation applied to graphene in the presence
of an arbitrary inhomogeneous electric and 
perpendicular magnetic field. It is shown that 
parabolic cylindrical functions, 
which have previously been used in literature, become 
inaccurate at high incident energies and low magnetic fields. A series expansion technique
is presented to circumvent this problem. An alternate method using asymptotic expressions is also
discussed and the relative merits of the two methods are compared. 
\end{abstract}

\vspace{2pc}
\maketitle

\section {Introduction}

Graphene's \cite{Geim2007, Geim2009} near perfect two-dimensional configuration and its unique 
electronic properties\cite{CastroNeto2009, Peres2010} have made it one of the widely studied materials in recent times. Its electrons have been found to obey a linear dispersion relation near the Fermi energy which makes them behave like
massless relativistic particles in two dimensions. As a result, they obey the massless Dirac equation instead of the Sch\"odinger equation. One of the important consequences of the relativistic behaviour of transport electrons is their inability to be confined
by an electrostatic barrier, a phenomenon known as Klein tunnelling \cite{Katsnelson2006}. The alternate strategy of confining these Dirac fermions using magnetic fields has been proposed\cite{DeMartino2007, DeMartino2007a}. Consequently, there has been a lot of interest in electron transport through magnetic barriers in graphene. Moreover, 
building functional electronic devices using graphene relies on being able to control the electronic transport by 
the application of electromagnetic fields. In this context, electron transmission through varying regularly and irregularly shaped barriers of both scalar and vector potentials becomes an important problem. Proper analysis of such barriers calls for the development of efficient numerical techniques.

In this work, we are interested in developing a general algorithm for the 
calculation of electron transmission in graphene through inhomogeneous
electric and magnetic fields. We only consider magnetic fields perpendicular to the  plane of the graphene sheet. We also restrict ourselves to the quasi one-dimensional problem which implies that the fields are invariant in the y-direction and the electronic plane wave is incident on it at an arbitrary angle.

From a mathematical viewpoint, this involves solving the massless 
Dirac equation which consists of two first-order coupled ordinary differential
equations with arbitrary values of electric and magnetic fields. We use the well-known transfer matrix method
to solve this problem. This method has previously been applied to problems in optics
\cite{Ghatak1987} and quantum mechanics \cite{Jonsson1990}. It is computationally
easy to implement, involving only the multiplication of $2 \times 2$ matrices. 
It has been used to study the scattering problem for the Schr\"odinger equation.\cite{Boonserm2009, Boonserm2010}.
The method has also been extended to solving any homogeneous ordinary linear
differential equation\cite{Khorasani2003b}.

Transfer matrix methods to solve the electron transport
problem in graphene have been studied extensively:
\cite{Katsnelson2006,  DeMartino2007} have applied it to single magnetic barriers;
it has been used in \cite{RamezaniMasir2008} to study the transmission
through multiple magnetic barriers in graphene;  in \cite{Barbier2009},
it is used for electrostatic barriers in  bilayer graphene;
in \cite{Masir2010, Li2010} for graphene superlattices; in \cite{Sun2010}  for fractally 
arranged magnetic barriers; and in \cite{RamezaniMasir2010} for tunnelling
through electric barriers in the presence of a magnetic field.

Although we proceed along similar lines, we show that the parabolic cylindrical functions (Weber functions) that have been used in literature can cause significant numerical difficulties  at low magnetic fields or at high incident energies 
and we use a series expansion to solve the differential equation
in order to avoid this problem. This forms the main result of this paper. 
Thus, we provide a uniform framework though this series expansion method and widen the applicability 
of the transfer matrix method for a large range of incident energies and magnetic fields.
Since the ballistic transport regime of Graphene based device is now being studied extensively both experimentally and theoretically \cite{YoungKim2011}, our scheme will be quite useful to understand some of such future experiments.
We also discuss an alternative method based on approximating Weber functions by their asymptotic form.
This method is applicable only within the asymptotic regime whereas the series method
is applicable to the entire range of magnetic fields and energies. We show that our method provides accurate results 
in this range also.

In the special case when the average length across 
which the vector potential varies is  
smaller that the typical magnetic length $l_B = \sqrt{({\hbar c})/({eB})}$, 
the magnetic barrier can be approximated by a delta function.
Analytic solutions for magnetic barriers modelled as 
a series of delta functions are well known\cite{Gumbs1985}, \cite{Ghosh2009}.
The transfer matrix method is more general and can be used even when this condition
doesn't hold.

Solving the Schr\"odinger or the Dirac equation includes two different kinds of problems: the eigenvalue problem and the scattering problem.
The eigenvalue problem involves finding the energy eigenvalues of the Hamiltonian and is used to find the allowed energy levels of bound states. 
The scattering problem, which is the one we tackle in this paper, involves the calculation of the transmission and reflection coefficients,
formally defined as the ratio of the flux of particles transmitted or reflected from a potential configuration to 
the flux that is incident on it. It leads to a second order homogeneous differential equation, 
the ubiquitous wave equation, which in one dimension is given by:
\begin{equation}
\label{eq:waveeqn}
 \psi''(x) + k^2(x)\psi(x) = 0 \quad (k\in \bf{C})
\end{equation}
The transfer matrix method involves division of the one-dimensional domain into slices and taking an appropriate approximation of $k^2(x)$ in each
slice. The equation for each slice is then solved and the continuity conditions are used at the interfaces of two such slices. The exact solution of the equation
in each slice depends on the form of $k^2(x)$ chosen. For example, for the Schr\"odinger equation, a piecewise constant approximation of $k^2(x)$ leads
to complex exponential solutions in each slice and a piecewise linear approximation leads to a solution basis consisting of the Airy functions\cite{Lui1986}.

In the case of graphene, we consider both scalar potentials (electrostatic fields) and vector potentials (magnetic fields), which lead to a piecewise linear vector potential and a piecewise constant scalar potential. The form of the resulting equation 
is:
\begin{equation}
\label{eq:eq_basic}
 \psi(x)'' + \left[ \alpha^2 - p  - (\beta + px)^2 \right]\psi(x) = 0
\end{equation}
where $\alpha, \beta, p \in \bf{R}$ and are explained later in detail. This equation admits parabolic cylindrical functions as the solution basis. We show that using these becomes computationally infeasible as $p\rightarrow0$
which corresponds to low magnetic fields and therefore an alternate solution basis is called for. We obtain this using the Frobenius method and find basis functions that tend to complex exponentials as $p\rightarrow0$. 

It is also necessary to restrict the transfer matrix method to cases where the 
magnetic field is non-zero only over some closed bounded (compact) interval.
This divides space into three regions and the solution in the first and last regions are complex
exponentials representing incoming and outgoing waves. From a physical point of view,
this condition is necessary because if $\forall x, B\neq0$, such as with a uniform 
magnetic field, the wavefunction gets localized along the spatial direction $x$.

In Section~\ref{sec:goveq} are outlined the equations to be solved and the 
notation used. In Section~\ref{sec:tmat}, the transfer matrix method is discussed.
In Section~\ref{sec:solution}, methods are outlined to solve Equation~\ref{eq:eq_basic}:
Section~\ref{subsec:pbdv} details the method previously found in literature along 
with its limitations. Section~\ref{subsec:asy} is a method based on asymptotic expansions,
and Section~\ref{subsec:frob} is the proposed alternative series method. Finally, in Section~\ref{sec:results},
we apply this method to a number of cases and present the results obtained.

\begin{figure}
 \label{fig:slices}
\centering
    \includegraphics{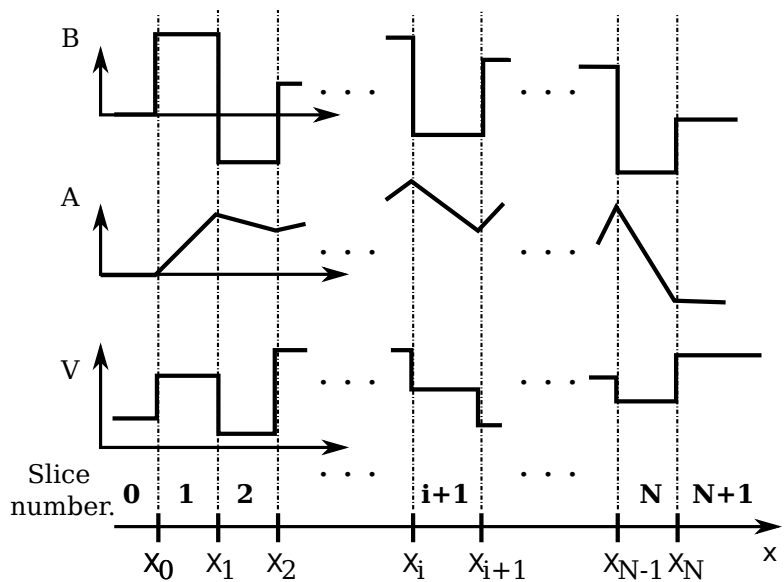}
\caption{Piecewise constant scalar potential and piecewise linear vector potential
showing the notation for x coordinates and slices used in the paper}
\end{figure}

\section{Governing Equations} \label{sec:goveq}

The governing massless Dirac equation is given by $ H\psi = E\psi$ 
where the Hamiltonian is given by 
\begin{equation}
\label{eq:hamiltonian}
 H = v_f \ \vec\sigma.(\vec{p} + e\vec A(x)) + V(x)
\end{equation}
and $\psi=[\psi_1, \psi_2]^T$ is the two component wavefunction,
$\vec{\sigma} = \sigma_x\hat{i}+\sigma_y\hat{j}$ with
$\sigma_{x,y}$ denoting the Pauli spin matrices.

Both the magnetic field $B$ and scalar potentials $V$ are discretized
and the magnetic field $B$ is converted to vector potential $A$ in the Landau gauge. 
The discretisation scheme that we have used is shown in Figure~\ref{fig:slices}.  
Slices are numbered from $0$ to $N+1$, with the leftmost and rightmost
slices unbounded. The boundaries between slices are denoted by $x_i$ with  
 $1 \le i \le N$. Therefore the i\textsuperscript{th} slice is bounded by 
$x= x_{i-1}$ and $x = x_i$. We denote the magnetic field,
scalar potential, and y component of the vector
potential in slice $i$  by the notation
 $B_i$, $V_i$, $A_i$ respectively.

For well defined transmission and reflection coefficients,
it is necessary to have zero magnetic field in the first and last slice, $B_0 = B_{N+1} = 0$, so that 
the solution can be expressed as complex exponentials which represent incoming 
and outgoing plane waves. The only non-zero component of $\vec{A}$ 
is $A_y(x)$ and is denoted by $A$. The functional form of the vector 
potential in the i\textsuperscript{th} slice is  
$A_i = C_i + B_i (x-x_{i-1})$ 
where $x_{i-1}$ represents the left edge of the i\textsuperscript{th} slice,
with $x_{-1}$ any conveniently chosen value (because $B_0=0$), and 
$ C_i=\sum_{j=0}^{i-1} B_j (x_{j} - x_{j-1})$, $C_0=0$.

The equations given above are converted to dimensionless form by defining two new variables.
We substitute $x' = x/x_s $ and $A' = A/A_s$. These can be also be thought of as scaling 
factors and as we shall see later, their exact values are important in computations. In terms of these 
scaled units, $A'_i = c_i + b_i (x' - \delta_{i-1}) $. It can immediately be seen that 
$b_i = B_i x_s / A_s$, $c_i = C_i/A_s$ and $\delta_{i-1} = x_{i-1}/ x_s$.
In terms of individual components and scaled units, Equation~\ref{eq:hamiltonian} is: 
\begin{eqnarray*}
 \frac{-i}{x_s}\frac{\partial \psi_2}{\partial x'}  - ik_y\psi_2  + i\frac{e}{\hbar}(-A_s A'\psi_2)  = (\epsilon - \tilde{v})\psi_1
\\
 \frac{-i}{x_s}\frac{\partial \psi_1}{\partial x'}  + ik_y\psi_1  + i\frac{e}{\hbar}(A_s A'\psi_1)  =  (\epsilon - \tilde{v})\psi_2
\end{eqnarray*}
where  
$\epsilon = E/\hbar v_f$ and $\tilde{v} = V/\hbar v_f$ (these have the units of $[L]^{-1}$).
The y-invariance of the problem leads to $\frac{\partial}{\partial y} = ik_y$ where $k_y = \epsilon \sin(\phi)$ with $\phi$
being the angle of incidence. We seek the transmission as a function of $\phi$.
The equations are decoupled bearing in mind 
that $\tilde{v}$ is constant in each slice and 
$A_i' = c_i + b_i (x - \delta_{i-1})$ is a function of $x$.
$\psi_1$ and $\psi_2$ are then governed by the following relations:
\begin{equation}
 \label{eq:main1}
 \frac{\partial^2 \psi_1}{\partial x'^2} + \left[ x_s^2(\epsilon - \tilde{v})^2 - \frac{e}{\hbar}x_s A_s \frac{\partial A'}{\partial x'} - x_s^2\left(k_y + \frac{e}{\hbar}A_s A'\right)^2\right] \psi_1 = 0
\end{equation}
\begin{equation}
 \label{eq:main2}
\psi_2 = \frac{1}{(\epsilon - \tilde{v})} \left[\frac{-i}{x_s}\frac{\partial \psi_1}{\partial x'} + ik_y\psi_1 + i\frac{e}{\hbar}(A_s A'\psi_1)\right]
\end{equation}
We use the standard technique of calculating $\psi_1$ from Equation~\ref{eq:main1} and 
calculating $\psi_2$ by backsubstituting $\psi_1$ in Equation~\ref{eq:main2}. In Section~\ref{sec:solution}, 
these equations are solved for a particular slice and in Section~\ref{sec:tmat}, these solutions
are used to construct the transfer matrix and completely solve the transmission problem.

\section {The transfer matrix method} \label{sec:tmat}
The transfer matrix method relies on the availability of two linearly 
independent analytic solutions of Equation~\ref{eq:main1} and Equation~\ref{eq:main2}.
If the two  linearly independent solutions of $\psi_1$ are denoted by $\psi_1^A$ and $\psi_1^B$, 
and the corresponding solutions for $\psi_2$ are $\psi_2^A$ and $\psi_2^B$, 
the transfer matrix denoted by $M_i$ is
such that the solution of the i\textsuperscript{th} slice is given by:
\begin{equation}
\label{eq:matrixeq1}
 	  \left [ \begin{array}{c}
	           \psi_1 \\ \psi_2 \\
	          \end{array}
	  \right] = 
	  M_i \left [ \begin{array}{c}
	           A_i \\ B_i \\
	          \end{array}
	  \right]
	,
	M(x) = \left[ \begin{array}{cc} \psi_1^A(x) & \psi_1^B(x) \\ \psi_2^A(x) & \psi_2^B(x) \end{array} \right] 
\end{equation}
From the continuity of $\psi_1$ and $\psi_{2}$ across the boundaries, we have
\begin{equation}
 M_i(x_i)\left [ \begin{array}{c} A_i \\ B_i \\ \end{array} \right] = 
M_{i+1}(x_i)\left [ \begin{array}{c} A_{i+1} \\ B_{i+1} \\ \end{array} \right] 
\quad \forall \quad 0 \le i \le N
\end{equation}
This allows us to formulate a recurrence relation between the coefficients $A_i,B_i$ and $A_{i+1}, B_{i+1}$.
Continuing in a similar manner, we relate $A_0, B_0$ with $A_{N+1}, B_{N+1}$ which then gives us 
the reflection and transmission coefficients.
\begin{equation}
\label{eq:fulltransfermatrix}
\left [ \begin{array}{c} A_0 \\ B_0 \\ \end{array} \right] = P \left [ \begin{array}{c} A_{N+1} \\ B_{N+1} \\ \end{array} \right] 
,
P = \prod_{j=0}^{N}M_{j}(x_j)^{-1} M_{j+1}(x_j) 
\end{equation}

We refer to the expression $M_j(x_{j-1}) M_j^{-1}(x_j)$ occurring in the expression for $P$  as the 
transfer matrix for the j\textsuperscript{th} slice. It can be easily proven that this term is 
independent of the basis functions chosen in that slice.

The expression for the transfer matrix given in Equation~\ref{eq:fulltransfermatrix}
can usually be simplified if the solution in each slice can be solved 
in a local coordinate system with its origin on the left edge
of that slice. This can always be done by shifting the origin in the wave equation, Equation~\ref{eq:waveeqn}. Then
the matrix  $M(x)$ depends only on $x-x_{i-1}$.
Thus, if $M(x) = N(x - x_{i-1})$, substitution in Equation~\ref{eq:fulltransfermatrix} gives this formula:
\begin{equation}
\label{eq:tmat1}
P = \prod_{j=0}^{N}N_{j}(x_j - x_{j-1})^{-1} N_{j+1}(0) 
\end{equation}
where $x_{-1}$ is a suitably chosen constant as explained earlier. We have used this expression
in our computations.

\subsection{Form of incident, transmitted and reflected waves}
In the first and last region, the magnetic field is chosen to be zero so that the solution reduces to complex 
exponentials of the form $\exp(\pm ikx)$ that represent the incident, reflected and transmitted waves.

In contrast to the Schr\"odinger equation in which $\exp(+ikx)$ represents right propagating waves,
and $\exp(-ikx)$ represents left propating waves, in the case of the Dirac equation $\psi_1 = \exp(+ikx)$ 
may represent either right or left propagating waves. If, in a slice, $E>V$, the probability flux corresponding
to $\psi_1 = \exp(+ikx)$ is positive implying that the wave is right propagating. On the other hand, if $E<V$, 
the flux corrresponding to the same wavefunction is negative and so it represents a left propagating wave. 

The incident wave, reflected wave and transmitted wave
are given by $\exp({s_0ik_0x})$, $r\exp({-s_0ik_0x})$ and $t\exp({s_{N+1}ik_{N+1}x})$ respectively where $s_i = \textrm{sign}(E-V_i)$
and $k_{j} = x_s \sqrt{ (\epsilon - \tilde{v_j})^2 - (k_y + \frac{e}{\hbar}A_s c_j)^2}$. The corresponding probability 
currents in the x-direction, within a constant, given by $\psi^\dagger\sigma_x\psi$ are 
$J_i=2k_0/|\epsilon - \tilde{v}_0|$, $J_r=-2k_0|r|^2/|\epsilon - \tilde{v}_0|$ and $J_t = 2k_{N+1}|t|^2/|\epsilon - \tilde{v}_{N+1}|$.
The transmission and reflection coefficients are given by:
\begin{eqnarray*}
\label{eq:rt}
R &= -J_r/J_i =  |r|^2 \\ 
T &= J_t/J_i  =  |t|^2 \frac{ k_{N+1}/|\epsilon - \tilde{v}_{N+1}|}{k_{0}/|\epsilon - \tilde{v}_0|}
\end{eqnarray*}
If, however $k_{N+1}$ is imaginary, $J_t=0$ and hence $T=0$.

The elements of the transfer matrix $P$ (Equation~\ref{eq:tmat1}) relates the coefficients of the complex
exponentials with a positive or negative sign in the first layer to those in the last layer (denoted by
$e_{first}\pm$ and $e_{last}\pm$):
\begin{eqnarray*}
\left[ \begin{array}{c}e_{first}+ \\e_{first}-\end{array}\right] = P \left[ \begin{array}{c}e_{last}+ \\e_{last}-\end{array}\right]
&, \quad P= \left[\begin{array}{cc} a & b\\c & d\end{array}\right] 
\end{eqnarray*}

The values of $r$ and $t$ to be used in Equation~\ref{eq:rt} depend on the form that the incident, reflected 
and transmitted waves have; i.e., whether they are represented by complex exponentials with positive or negative signs.
The results are summarised in the following table:
\begin{equation}
\begin{array}{llcll}
E>V_0, & E>V_{N+1} &:&t=1/a & r=c/a\\
E>V_0, & E<V_{N+1} &:& t=1/b & r=d/b\\
E<V_0, & E>V_{N+1} &:& t=1/c & r=a/c\\
E<V_0, & E<V_{N+1} &:& t=1/d & r=b/d\\
\end{array}
\end{equation}

\section{Solving the Governing Equations}
\label{sec:solution}
We now solve equations~\ref{eq:main1} and \ref{eq:main2}
and find solution bases $\psi_1^{A,B}, \psi_2^{A,B}$  to construct the transfer matrix used in Equation~\ref{eq:tmat1}. To this end,
we introduce another change of variable with 
$ x'' = x' - \delta$ representing a translation 
of the origin to the left boundary of each slice.
For notational convenience, subscripts indicating the slice number
are omitted in this section. Defining dimensionless constants $ \alpha =  x_s (\epsilon - \tilde{v}) $, 
$p = \frac{e}{\hbar} x_s A_s b$ and $ \beta  =  x_s (k_y + \frac{e}{\hbar}A_s c)$,
equations~\ref{eq:main1} and \ref{eq:main2}
can be written in the dimensionless form
\begin{equation}
\label{eq:main3}
 \frac{d^2 \psi_1}{dx''^2 } + \left[ \alpha^2 - p  - (\beta + px'')^2 \right]\psi_1 = 0
\end{equation}
\begin{equation}
\label{eq:main4}
 \psi_2 = \frac{i}{\alpha} \left[ \frac{\partial \psi_1}{\partial x''} + \psi_1 ( \beta + p x'')\right]
\end{equation}

\subsection{Parabolic Cylindrical function solution} \label{subsec:pbdv}
We first discuss the well-known technique of using parabolic
cylindrical functions~\cite{Sun2010},\cite{DeMartino2007},\cite{Oroszlany2008}  to solve Equation~\ref{eq:main3} and \ref{eq:main4}.
The parabolic cylindrical equation in standard form is 
\begin{equation}
 \frac{\partial^2\psi}{\partial x^2} - \psi\left[ \frac{x^2}{4} + a\right] = 0
\end{equation}
Following the notation used by \cite{Zhang1996}, the two linearly independent 
solutions to the equation are given by $U(a,x) = D_\nu(x)$ and $V(a,x) = V_\nu(x)$
with $\nu = -(1/2 + a)$. Alternatively,
$D_\nu(x)$ and $D_\nu(-x)$ can also be used as linearly independent solutions.

For solving Equation~\ref{eq:main3}, three cases of $p>0$, $p<0$ and $p=0$ (corresponding
to positive, negative and zero magnetic field) need to 
be dealt with separately. When $p=0$, the solutions are complex 
exponentials:
\begin{equation}
\label{eq:complexexp}
 \psi_1 = e^{\pm i\sqrt{\alpha^2 - \beta^2}x''}
\\
 \psi_2 = \frac{1}{\alpha} \left[ \pm \sqrt{\alpha^2 - \beta^2} + i\beta \right] e^{\pm i\sqrt{\alpha^2 - \beta^2}x''}
\end{equation}
When $p \neq 0 $, the equation can be converted to standard form by substituting
$z = \sqrt{{2}/{|p|}} (\beta + p x'') $. The solutions are
$ \psi_1 = D_\nu(z), V_\nu(z), D_\nu(-z) $,
where either $D_\nu(z)$, $V_\nu(z)$ or $D_\nu(z)$, $D_\nu(-z)$ can be used;
$\nu$ is given by:
\begin{equation}
\label{eq:nu}
 \nu = 
\left\{ 
\begin{array}{ll}
 \frac{\alpha^2}{2p} - 1 & p>0\\
\frac{\alpha^2}{2|p|}  & p<0 
\end{array}
\right.
\end{equation}
The corresponding expression for $\psi_2$ given by Equation~\ref{eq:main4} is:
\begin{equation}
 \psi_2 = \frac{i}{\alpha} 
\left[ 
     -\sqrt{2|p|} {sign}(p)\frac{\partial \psi_1}{\partial z}  + \psi_1 z \sqrt{\frac{|p|}{2}}
\right]
\end{equation}
This can be further simplified by using standard recurrence relations relating
$D_\nu(z)$ and $V_\nu(z)$ to their derivatives and the simplified expressions are:
\begin{equation} 
\label{eq:psi1psi2exp}
 \begin{array}{rll}
     &  \quad \psi_1= \quad & \quad \psi_2= \quad\\ \\
p> 0:\quad & D_\nu(z) & \frac{i}{\alpha}\sqrt{2p}D_{\nu+1}(z) \\
           & D_\nu(-z) & \frac{-i}{\alpha}\sqrt{2p}D_{\nu+1}(-z) \\
           & V_\nu(z) & \frac{i}{\alpha}\sqrt{2p}(\nu+1) V_{\nu+1}(z) \\ \\
p<0: \quad & D_\nu(z) & \frac{i}{\alpha}\sqrt{2|p|}(\nu) D_{\nu-1}(z) \\
           & D_\nu(-z) & \frac{-i}{\alpha}\sqrt{2|p|}(\nu) D_{\nu-1}(-z) \\
           & V_\nu(z) & \frac{i}{\alpha}\sqrt{2|p|} V_{\nu-1}(z) 
 \end{array}
\end{equation}

We now discuss the limitations of this method. The function
$D_\nu(z)$ has a power-law dependence with $\nu$ and increases at a near-exponential rate with an increase in $\nu$ and reaches
$10^{308}$ at around $\nu=300$ which is the maximum representable 
double precision value on a computer.
It can be seen from Equation~\ref{eq:nu} that the parameter $\nu$ contains the term:
\begin{equation} 
\frac{\alpha^2}{2|p|} = \frac{x_s(\epsilon - \tilde{v})^2}{2\frac{e}{\hbar}A_s b}
= \frac{(\epsilon - \tilde{v})^2}{2\frac{e}{\hbar} B} 
\end{equation}
where the relation
 $b = B\frac{x_s}{A_s}$ has been used. From this, it can immediately  
be seen that $\nu$ increases with an increase in the incident
energy $\epsilon$ and increases with a decrease in the 
magnetic field $B$. Furthermore, the 
expression for $\nu$ is independent of any normalization or scaling factors.

The first problem with the parabolic
cylindrical function method is obvious: as the magnetic 
field decreases, $\nu$ gets larger and $D_\nu(z)$ becomes too large 
to be calculated in double precision. For an incident energy of 
82 meV (corresponding to a Fermi wavelength $k_f = 2\pi/\epsilon$ of 50 nm),
the minimum allowable magnetic field before this occurs is 0.017 T (corresponding to $\nu=300$). This makes it impossible to 
observe a transition between zero magnetic field and a finite 
magnetic barrier.  

Secondly, we are limited by the accuracy to which parabolic cylindrical functions
themselves are computed. Using the Fortran codes given in \cite{Zhang1996}, the
lowest magnetic field at which errors start showing up can be as high
as 0.6 T. These errors manifest themselves as unphysical results like 
abrupt discontinuities in the transmission plots. The transfer matrix can also
become near singular making it difficult to invert. In this case, we calculate the
pseudoinverse using singular value decomposition. We have verified that
round-off errors are the source of the problem by calculating the parabolic 
cylindrical functions in several ways, including one in which calculations are performed
in arbitrary precision before the result is rounded off to double-precision. Using 
this, we could go closer to the theoretical limit at $\nu=300$ mentioned above.

Examining calculations in literature using this method, we find that in most of the cases 
authors have limited their calculations to incident energies and magnetic field values
that result in small values of $\nu$. In \cite{Sun2010},
$\nu = 12.5$ has been used and \cite{DeMartino2007} have
used $\nu = 6.8$. This gives a rough indication of the range
in which parabolic cylindrical functions work.

\subsection{Asymptotic solution} \label{subsec:asy}

When the magnetic field is small, the parameter $\nu$ in $D_\nu(z)$ becomes large and using an 
asymptotic expansion instead of the parabolic cylindrical function is a possibility. 
This can be achieved using an asymptotic form for $D_\nu(z)$ for large $|\nu|$ which
 can be expressed as a product of a $\nu$-dependent term, $h(\nu)$, that causes exponential growth
of the function and some other factor. This large $\nu$-dependent term need not be explicitly computed 
because upon substitution in the expressions for the transfer matrix for the j\textsuperscript{th} slice 
given by $M_j(x_{j-1}) M_j^{-1}(x_j)$ in Equation~\ref{eq:fulltransfermatrix}, it gets cancelled out.
Asymptotic expansions that satisfy these criteria are available in \cite{Temme2000, Temme2001}. 	
Different expressions are applicable in different regions of the $\nu$-$z$ plane.

To elaborate, suppose there is a positive magnetic field in the j\textsuperscript{th} slice and transfer matrix for that slice is (see Equation~\ref{eq:psi1psi2exp})
\begin{equation} 
M_j(x) = \left[
 \begin{array}{cc}
  D_\nu(z(x)) & D_\nu(-z(x)) \\
  \frac{i}{\alpha}\sqrt{2p}D_{\nu+1}(z(x)) & \frac{-i}{\alpha}\sqrt{2p}D_{\nu+1}(-z(x)) \\
 \end{array} \right] 
\end{equation}
We now use the recurrence relation $D_{\nu+1}(z) = \frac{1}{2}z D\nu(z) - D'_\nu(z)$ and then substitute the asymptotic forms from 
\cite{Temme2000, Temme2001}. The factor of $h(\nu)$ can be factored out and gets cancelled. Similarly, when the magnetic field is
negative, the recurrence relation to be used is $D_{\nu-1}(z) = (\frac{1}{2}z D\nu(z) + D'_\nu(z))/\nu$

One of the limitations of this method is that it doesn't work at the turning points of the parabolic cylindrical differential equation
$ z = \pm 2 \sqrt{\nu + 1/2} $ when $\nu> - 1/2 $ and gives inaccurate answers close to those points. The other is that these
expressions work only in the asymptotic regime and not for all values of magnetic field and incident energy.

This scheme is similar to that used in \cite{DellAnna2009, DellAnna2011} where an asymptotic form
proposed in\cite{Watson1910, Schwid1935} has been used. However, the expression they have used is valid for $\nu\rightarrow\infty$
and $z \rightarrow 0$ with $z\sqrt\nu$ finite. This, however, will not work for a general case where we need an asymptotic form
that works for large $\nu$ and all $z$.

\subsection{Series solution} \label{subsec:frob}
Equation~\ref{eq:main3} 
predicts that the solutions with magnetic field ($p \ne 0$)
should smoothly tend to the solutions without magnetic
field ($p=0$). All the problems with the parabolic cylindrical functions method stem from our choice of basis functions
$D_\nu, V_\nu$ that do not tend to complex exponentials as $p \rightarrow 0$.
This leads us to choose solutions of Equation~\ref{eq:main3}
that do tend to complex exponentials as $p \rightarrow 0$.
These are discussed in this section.

This method to solve the equation relies on 
the Frobenius method which yields two linearly independent solutions of the form 
$\sum_0^{\infty} q_n (x'')^n$. With $\theta = -(\alpha^2 - \beta^2) + p$ and $\phi = 2\beta p$.
The coefficients $q_i$ for the two solutions $\phi_1$ and $\phi_2$ are given by
\begin{equation}
\begin{array}{lcccc}
\phi_1:&  q_0 = 1  &
  q_1 = 0 &
  q_2 = \frac{\theta}{2!} &
  q_3  = \frac{\phi}{3!} 
\end{array}
\end{equation}
\begin{equation}
\begin{array}{lcccc}
\phi_2: &
  q_0 = 0  &
  q_1 = 1 &
  q_2 = 0 &
  q_3 = \frac{\theta}{3!} 
\end{array}
\end{equation}
and for $n \ge 2$ given by the recurrence relation: 
\begin{equation}
  n \ge 2: q_{n+2} = \frac{\theta}{(n+1)(n+2)}q_n + \frac{\phi}{(n+1)(n+2)}q_{n-1} + \frac{p^2}{(n+1)(n+2)}q_{n-2}
\end{equation}
It can be shown that when $p=0$, the solutions tend to sine and cosine series.
When $p=0$ and $\beta=0$, the first
solution is $\cos(\alpha x'')$ and the second one is $\sin(\alpha x'')/\alpha$.
Similar results holds for when $p=0$ and $\beta \ne 0$. Then the first solution is
$\cos(\sqrt{\alpha^2 - \beta^2} x'')$ and the second one is 
$\sin(\sqrt{\alpha^2 - \beta^2} x''	)/\sqrt{\alpha^2 - \beta^2}$.
We choose the two linearly independent solutions 
to be used in the transfer matrix equation, Equation~\ref{eq:tmat1}
as $\psi_1^A=\phi_1 + ik\phi_2$ and $\psi_1^B=\phi_1 - ik\phi_2$ where $k=\sqrt{\alpha^2 - \beta^2 + p}$
because they reduce to complex exponentials in the limit $p\rightarrow 0$. In this way, the three cases of 
$p=0$, $p<0$ and $p>0$ do not need to be treated separately. Furthermore, Fuchs's Theorem\cite{Arfken2001} guarantees convergence
of the series solution.

Some care needs to be taken while summing up these series term by term. Under
usual circumstances, sine and cosine series are not directly summed up
because the terms increase before they start decreasing\cite{Press1992}.
However, for small arguments, the convergence is quick and manual summation becomes feasible.\
In the series that we have used, summation is possible only if 
$\alpha$, $\beta$, $p$ and $x''$ are small. We choose the scaling factors
$x_s$ and $A_s$ judiciously to make this possible.
This is critical to the process of manual summation.
By making $x_s$ small, the variables
$\alpha$, $\beta$ and $p$ can be made as small as desired.
However, $x'' = (x_i - x_{i-1})/x_s$ and decreasing $x_s$ increases $x''$. To avoid this, slices that are
very wide will sometimes need to be subdivided into narrower slices. In case any of 
these parameters are chosen incorrectly, the coefficients $q_i$ overflow or underflow
which can be detected quite easily. Also $x_{-1}$ should be taken to be equal to $x_0$.

It should be noted that the series needs to be summed up only once for each slice. 
Equation \ref{eq:tmat1} requires the evaluation of the series
at $x=0$ which doesn't require a series summation.
Calculation of $\psi_2$ is done using Equation~\ref{eq:main4}. This requires
evaluation of the derivative which can be easily done during the series summation.

\section{Numerical Examples} \label{sec:results}
\subsection{Implementation}
In the subsequent sections, we demonstrate the results of the numerical technique we have developed 
by applying it to a few specific cases. They include both cases with scalar potential only, vector potential only
and with both scalar and vector potentials.
The series summation algorithm was implemented in Fortran 95 and the rest of the program in python. Series summation
was performed till the error between partial sums was $10^{-20}$. 
The scale factor $x_s$ has been taken to be $10^{-8}$ nm in the results presented.

\subsection{Results for a single barrier}
\begin{figure}
\centering
 \label{fig:trans1}
    \includegraphics[scale=1]{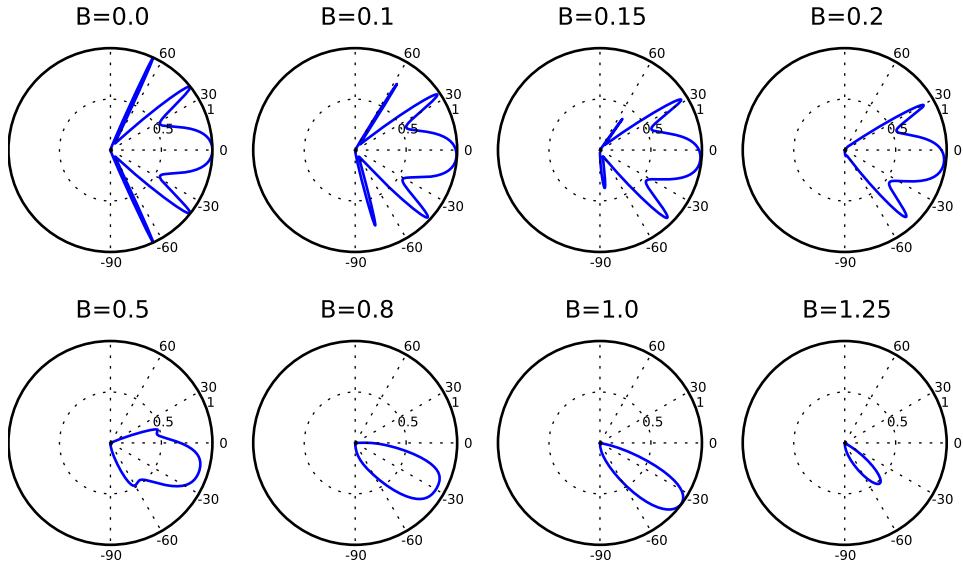}
\caption{Transmision through a single electrostatic barrier of width 100nm and height 180meV with a varying
 magnetic field across the 100nm barrier region. Incident energy is 82.66 meV}
\end{figure}

We consider an electrostatic potential barrier of width 100nm and height 180meV
and apply a varying magnetic field across this 100nm region. The incident energy 
chosen is around 82.66 meV corresponds to a Fermi wavelength of 50 nm.
The transmission plots using the series method are shown in Figure~\ref{fig:trans1}. 
The polar plots depict the transmission as a function of the angle of incidence.
The magnetic field values chosen for illustration are $0$ and between $0.1$($\nu = 72.08$) to $1.25$($\nu=5.76$)
A smooth transition from zero magnetic field to higher fields can be observed. The solutions 
calculated using the series method and parabolic cylindrical functions were found
to match at higher fields but the latter algorithm fails at low magnetic fields. 
So the low to high magnetic field transition cannot be observed by using parabolic
cylindrical functions. The asymptotic method solution matches with the results shown for 
low magnetic fields. For other combinations of incident energy and magnetic fields,
the asymptotic forumation can give incorrect results if the parabolic cylindrical functions
are calculated near the turning points. For comparison, similar plots using  the parabolic cylindrical functions
and asymptotic methods are shown in Figure~\ref{fig:trans1}.

\begin{figure}
\centering
 \label{fig:trans1B}
    \includegraphics[scale=1]{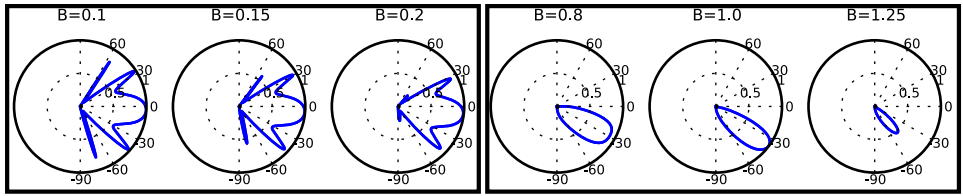}
\caption{Transmission through a single electrostatic barrier of width 100nm and height 180meV with a varying
 magnetic field across the 100nm barrier region. Incident energy is 82.66 meV. Left: 
Results calculated using asymptotic form of parabolic cylindrical functions valid at low magnetic fields. Right:
Results calculated using parabolic cylindirical functions valid at high magnetic fields.
Comparison with Fig. \ref{fig:trans1} shows that both in the high as well as low
magnetic field limit results can be reproduced by the current method
very accurately.}
\end{figure}

\subsection{Gaussian Barrier}
\begin{figure}
\centering
 \label{fig:trans3}
    \includegraphics[scale=0.9]{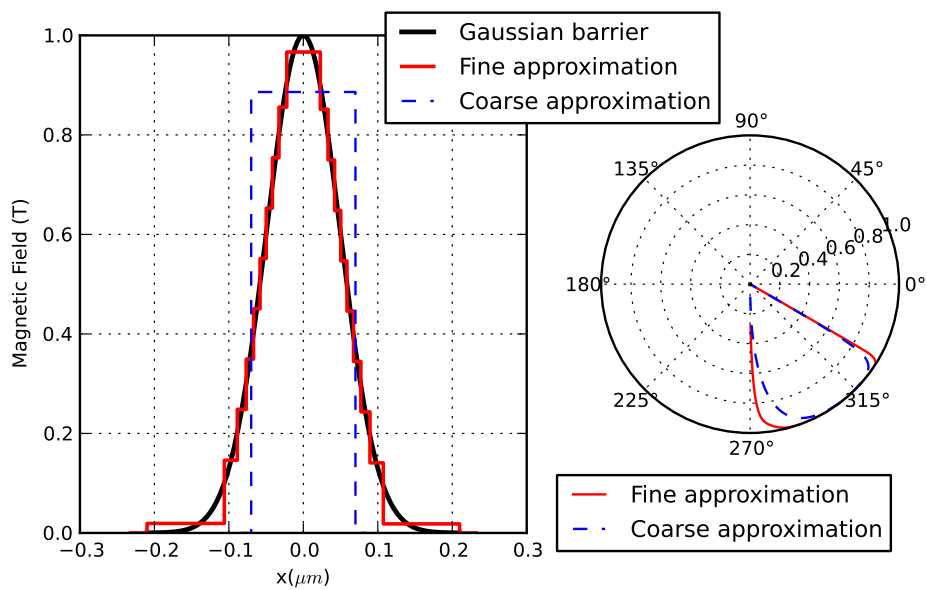}
\caption{A gaussian shaped magnetic barrier with two different approximations. 
(a) The gaussian curve with the coarse and fine approximations  (b) Transmission plots
for both approximations}
\end{figure}

We consider a single gaussian-shaped magnetic field barrier with no
electrostatic field and compute the transmission by using a coarse and a fine piecewise
approximation. In the coarse approximation, it is reduced to a a single square barrier and 
the fine approximation consists of it being approximated by as a series of barriers of varying height. 
The two approximations are shown in Figure~\ref{fig:trans3}.
The $1/e$ width of the gaussian curve is 140nm and the peak magnetic
field is 1T. The coarse approximation consists of a single barrier of width
140nm and magnetic field $\pi/2$ T. The fine approximation consists of
the division of the gaussian barrier into 21 slices with a maximum field 
variation of not more than 0.1T taken to be constant. The incident energy is 82.66 meV.
The two transmission plots are also shown. This example
clearly demonstrates the utility of being able to perform computations
for low magnetic fields even if the peak field value is high.

\subsection{Experimental Data}

\begin{figure}
 \centering
 \label{fig:yacoby}
    \includegraphics[scale=0.9]{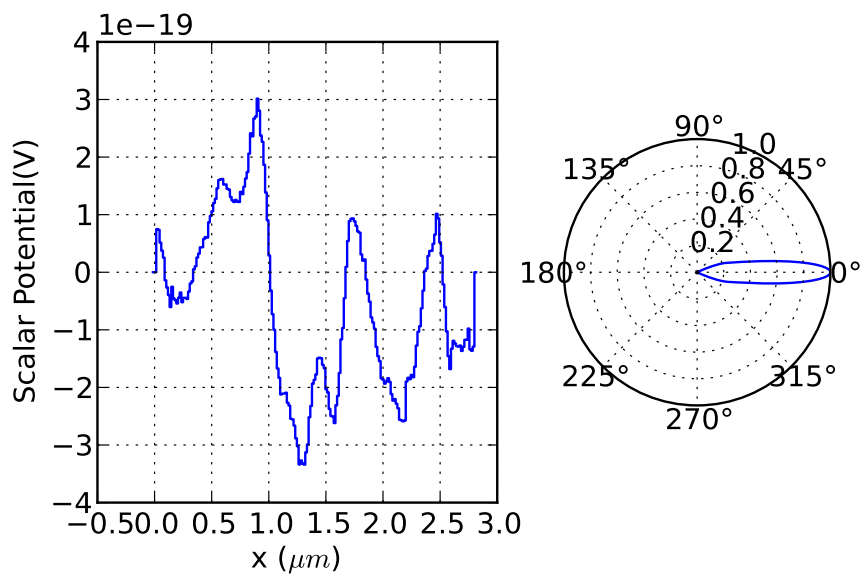}
\caption{Scalar potential distribution from experimental data 
given in \cite{Martin2007a} and the corresponding transmission.}
\end{figure}

We calculate the transmission in graphene based on the experimental data given 
in \cite{Martin2007a}. They have shown that the presence of disorder in graphene gives rise to 
localised charge distributions on the surface, or as they are called, electron and 
hole puddles. They have also measured this charge distribution. It is known that a scalar potential 
applied to graphene sheet shifts the Dirac point and leads to charge accumulation. We therefore
model the charge as arising from a scalar potential distribution proportional to it. We have
calculated the transmission corresponding to a one-dimensional potential extracted from their experimental data
scaled by an arbitrary factor of $10^{-29}$. The scalar potential and the transmission 
data are shown in Figure~\ref{fig:yacoby}.

\subsection{Random magnetic fields}

It has been shown that any elastic deformation in the graphene sheet manifests 
itself as effective gauge fields acting on charge carriers\cite{Guinea2009}. This can be caused by 
a corrugated substrate or by the intrinsic thermodynamic ripples in graphene. 
It has also been shown that several electronic devices can be built
by controlling the strain.\cite{Pereira2009}. 

The relation between a strain field and gauge fields is given in 
\cite{Guinea2009}. For a strain field with tensor components $u_{xx}$ and $u_{yy}$ denoting
the normal strain and $u_{xy}$ the shear strain, the relation to scalar and vector potentials
is as follows ($\beta, t, a, c, g$ are constants)

\begin{eqnarray*}
 A_x &= c \frac{\beta t}{a} (u_{xx} - u{yy}) \\  
 A_y &= - c \frac{\beta t}{a}  u_{xy} \\
 V & = g (u_{xx} + u{yy})
\end{eqnarray*}

Therefore, a strain field can be modelled as a gauge field. In the special case 
that only x-dependent shear strain is present, the only component of the equivalent
magnetic field is $A_y(x)$ which is a Landau gauge representable vector potential.

\begin{figure}
\centering
 \label{fig:trans2}
    \includegraphics[scale=1]{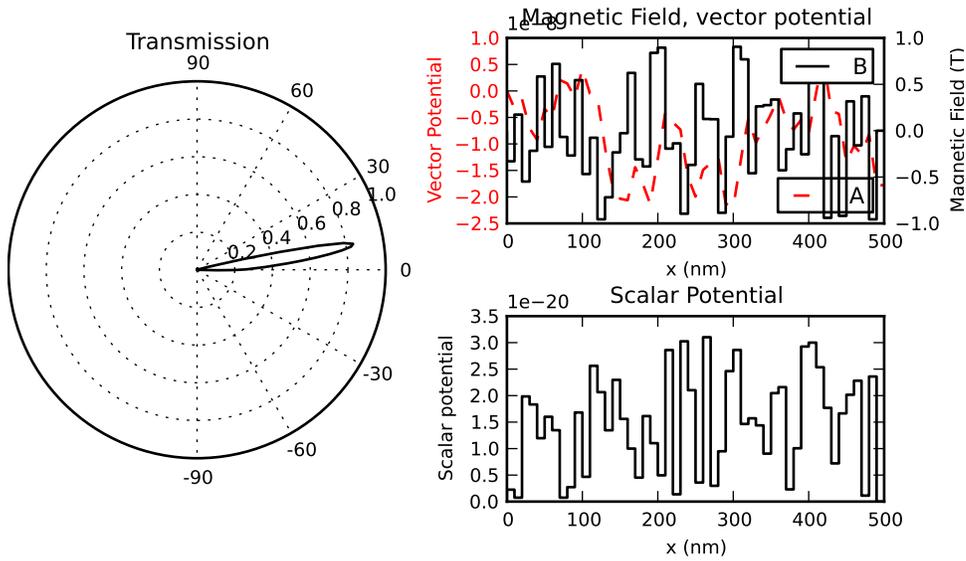}
\caption{Computation with random fields. (a) Transmission plot (b) Magnetic Field and vector potential
(c) Scalar potential}
\end{figure}

We carry out a transmission calculation in the presence of disorder
with the magnetic field and
scalar potential chosen randomly. 
Fifty slices, each 10 nm wide are taken with 
the magnetic field in each slice uniformly distributed between -1 T and 1 T.
The scalar potential in each slice is uniformly distributed between
0 and 200meV. A typical result is given in Figure~\ref{fig:trans2}, where the magnetic field, scalar and vector potentials are shown alongwith the resultant transmission.

\subsection{Application to bilayer graphene}
This series technique can also be extended to bilayer graphene in the presence of electrostatic and 
magnetic fields to obtain the transmission at high energies and low magnetic fields. This method 
has been used in a recent communication \cite{Agrawal2011}.

\section{Conclusion}
We have applied the transfer matrix method to solve transmission problems in graphene
in the presence of inhomogeneous electric and magnetic fields.
We have brought out some of the difficulties associated with parabolic cylindrical
functions and proposed a method to get around its limitations by changing the basis
functions to a series solution which tends to complex exponentials. Despite the overhead of 
numerically computing a series sum, our method is robust and easy to implement with different cases not needing 
separate treatment compared to the use of parabolic cylindrical functions
with or without asymptotic expansions. We also believe that the method is quite general and can be profitably 
employed whenever the wave equation is being solved with the transfer matrix method.

\section{Acknowledgments}
This work is supported by grant
SR/S2/CMP-0024/2009 by Science and Engineering Research Council, DST,
Govt. of India. 

\bibliographystyle{unsrt}

\bibliography{library}

\end{document}